\documentclass[12pt]{article}
 \topmargin{-2cm}
 \hoffset = - 0.5 in
 
\begin{document}
 \title{
 Stability of scaling regimes in $d\geq 2$  developed
  turbulence with weak anisotropy}

 \author{
M. Hnatich$^{1}$, E. Jonyova$^{1,2}$, M. Jurcisin$^{1,2}$,
 M. Stehlik$^{1}$       \\
$^{1}$ {\small Permanent address:} {\small{\it Institute of
Experimental Physics, SAS, Ko\v{s}ice, Slovakia}} , \\ $^{2}$
{\small{\it  Joint Institute for Nuclear Research,
 Dubna, Russia}}
 }
\date{}
\maketitle
\begin{abstract}
The fully developed turbulence with weak anisotropy is
investigated by means of renormalization group approach (RG) and
double expansion regularization for dimensions $d\ge 2$. Some
modification of the standard minimal substraction scheme
 has been used to analyze stability of the Kolmogorov scaling
regime which is governed by the renormalization group  fixed
point. This fixed point is unstable at $d=2$; thus, the
infinitesimally weak anisotropy destroyes above scaling regime in
two-dimensional space. The restoration of the stability of this
fixed point,  under transition from $d=2$ to $d=3,$ has been
demonstrated at borderline dimension $ 2<d_c<3$. The results are
in qualitative agreement with ones obtained recently in the
framework of the usual analytical regularization scheme.
\end{abstract}

\section{Introduction}

A traditional approach to the description of fully developed
 turbulence is
based on the stochastic Navier-Stokes equation \cite{wyld}.
The complexity of this equation leads to great difficulties
which do not allow one to solve it even in the simplest case
when one
assumes the  isotropy of the system under consideration. On
the other hand, the isotropic turbulence is almost delusion and if
exists is still rather rare. Therefore, if one wants to model more
or less realistic developed turbulence, one is pushed to consider
anisotropically forced turbulence rather than isotropic one. This,
of course, rapidly increases complexity of the corresponding
differential equation which itself has to involve the part
responsible for description of the anisotropy.
An exact solution of the stochastic Navier-Stokes equation
 does not exist and one is forced to find out some
convenient methods to touch the problem at least step by step.

A suitable and also powerful tool in the theory of developed
turbulence is the so-called renormalization group (RG)
method\footnote{ Here we consider the quantum-field
renormalization group approach \cite{AVP} rather than
the Wilson renormalization group
technique \cite{WIL} }. Over the last two decades the RG technique
has widely  been used in this field of science and gives answers to
some principal questions (e.g. the fundamental description of the
infra-red (IR) scale invariance) and is also useful for
calculations of many universal parameters (e.g. critical
dimensions of the fields and their gradiens etc.).
A detailed survey of this questions
can be found in Refs. \cite{ADZ1,VASB} and Refs. therein.

In early papers, the RG approach has been applied only to the
isotropic models of developed turbulence. However, the method can also
be used (with some modifications) in the theory of anisotropic
developed turbulence.
A crucial question immediately arises here, whether the principal
properties of the isotropic case and anisotropic one are the same at
least at the qualitative level. If they are, then it is possible to
consider the isotropic case as a first step in the investigation of
real systems.
On this way of transition from the isotropic developed turbulence
into the anisotropic one we have to learn whether the scaling regime
does remain stable under this transition.
That means, whether the stable fixed points of
the RG equations remain stable under the influence of anisotropy.

Over the last decade a few papers have appeared in which the above
question has been considered. In some cases it has been found out
that stability really takes place (see,e.g.\cite{RUB,KIM}). On the
other hand, existence of systems without such a stability has
been proved too.
As has been shown in Ref.\,\cite{ADZ2}, in the anisotropic
magnetohydrodynamic developed turbulence a stable regime generally
does not exist. In \cite{KIM,BUSA}, d-dimensional models
with $d>2$ were investigated for two cases: weak anisotropy
\cite{KIM} and strong one \cite{BUSA}, and it has been shown that the
stability of the isotropic fixed point is lost for dimensions
$d<d_c=2.68$. It has also been shown that stability of the fixed
point even for dimension $d=3$ takes place only for sufficiently
weak anisotropy.
The only
problem in these investigations is that it is impossible to use
them in the case $d=2$ because new ultra-violet (UV) divergences
appear in the Green functions when one considers $d=2$,
and they were not taken into account in \cite{KIM,BUSA}.

In \cite{HON}, a correct treatment of the two-dimensional
isotropic turbulence  has been given. The correctness in the
renormalization procedure has been reached by introduction into
the model a new local term (with a new coupling constant) which
allows one to remove additional UV-divergences. From this point of
view, the results obtained earlier for anisotropical developed
turbulence presented in \cite{OLLA} and based on the paper
\cite{RON} (the results of the last paper are in conflict with
\cite{HON}) cannot be considered as correct because they are
inconsistent with the basic requirement of the UV-renormalization,
namely with the requirement of the localness of the counterterms
\cite{COL,ZIN}.

The authors of the recent paper \cite{ANT} have used the
double-expansion procedure introduced in \cite{HON} (this
procedure is a combination of the well-known Wilson dimensional
regularization procedure and the analytical one) and the minimal
substraction (MS) scheme \cite{hoft} for investigation of
developed turbulence with weak anisotropy for $d=2$. In such a
perturbative approach the deviation of the spatial dimension from
$d=2$, $\delta=(d-2)/2$, and the deviation of the exponent of the
powerlike correlation function of random forcing from their
critical values, $\epsilon$, play the role of the expansion
parameters.

The main result of the paper was the conclusion that the
two-dimensional fixed point is not stable under weak anisotropy.
It means that 2d turbulence is very sensitive to the anisotropy
and no stable scaling regimes exist in this case.
In the case $d=3$, for both the isotropic turbulence and anisotropic one,
as it has been mentioned above,
existence of the stable fixed point, which  governs  the Kolmogorov
asymptotic regime, has been established by means of the RG approach
by using the analytical regularization procedure \cite{RUB,KIM,BUSA}.
One can make analytical continuation from $d=2$ to the
three-dimensional turbulence (in the same sense as in the theory
of critical phenomena) and verify
whether the stability of the fixed point (or, equivalently,
stability of the Kolmogorov scaling regime) is restored.
From the analysis made in Ref.\,\cite{ANT} it follows that it is
impossible to restore the   stable regime by transition from
dimension $d=2$ to $d=3$. We suppose  that the main reason for the above
described discrepancy is related to the straightforward
application of the standard MS scheme. In the standard MS scheme
one works  with the purely divergent part of the Green functions only,
and in concrete calculations its dependence on the space dimension
$d$ resulting from the tensor nature of these Green functions is
neglected (see Sect.3).
In the case of isotropic models, the stability of the fixed points
is independent of dimension $d.$
However, in anisotropic models the stability of fixed points
depends on the dimension $d$, and consideration of the tensor
structure of Feynman graphs in the analysis of their divergences
becomes important.

In present paper we suggest applying modified MS scheme in which
we keep the $d-$dependence of UV-divergences of graphs.
We affirm that after such a
modification the $d$-dependence is correctly taken into account and
can be used in investigation of whether it is possible to restore the
stability of the anisotropic developed turbulence for some
dimension $d_c$ when going from a two-dimensional system to
a three-dimensional one.
In the limit of infinitesimally weak anisotropy for the physically
most reasonable value of $\epsilon=2$, the value of the borderline
dimension is $d_c=2.44.$
Below  the borderline dimension, the stable regime of the fixed point of
the isotropic developed turbulence is lost by influence of weak
anisotropy.

It has to be mentioned that a similar idea of "geometric factor"
was used in Ref.\,\cite{frey} in RG analysis of the
Burgers-Kardar-Parisi-Zhang equation but the reason to keep
the $d-$dependence of divergent parts of the graphs was to take
correctly into
account the finite part of one-loop Feynman diagrams in the
two-loop approximation.
In the present paper, we shall not discuss it in detail because
the critical analysis of the results obtained in \cite{frey} was given
in \cite{wiese}.

The paper is organized as follows. In Section 2 we give the
quantum field functional formulation of the problem of the fully
developed turbulence with weak anisotropy. The RG analysis is
given in  Section 3 when we discuss the stability of the  fixed
point obtained  under weak anisotropy. In Section 4 we discuss our
results. Appendix I contains expressions for the divergent parts of
the important graphs. At the end, Appendix II contains analytical
expressions for the fixed point and the equation which describes its
stability in the limit of the weak anisotropy.

\section{Description of the Model. UV-divergencies}

In this section we give the description of the model. As has already
been discussed in the previous section, we work with fully developed
turbulence and assume weak anisotropy of the system. It means that
the parameters that describe deviations from the fully isotropic case
are sufficiently small and allow one to forget about corrections
of higher degrees (than linear) which are made by them.

In the statistical theory of anisotropically developed turbulence the
turbulent flow can be described by a random velocity field
$\vec v (\vec x ,t)$
and its evolution is given by the randomly forced Navier-Stokes equation
\begin{equation}
\frac{\partial \vec v}{\partial t} +
({\vec v} \cdot \vec \nabla)\vec v
-\nu_0 \Delta \vec v - \vec f^A = \vec f, \label{NS}
\end{equation}
where we assume incompressibility of the fluid, which is
mathematically given by the well-known conditions $\vec\nabla
\cdot \vec v =0$ and $\vec\nabla\cdot\vec f =0$. In eq.(\ref{NS})
the parameter $\nu_0$ is the kinematic viscosity
(hereafter all parameters
with subscript $0$ denote bare parameters of unrenormalized theory,
see below), the term $\vec f^A$ is
related to anisotropy and
will be specified later.  The large-scale random force per  unit
mass $\vec f$ is assumed to have Gaussian statistics defined by
the averages
\begin{eqnarray}
\langle f_i \rangle =0,\,\, \langle f_i(\vec x_1 ,t)f_j(\vec x_2 ,t)
\rangle=D_{ij}(\vec x_1 - \vec x_2 ,t_1-t_2).
\end{eqnarray}
The two-point correlation matrix
\begin{eqnarray}
D_{ij}(\vec x ,t)=\delta (t) \int \frac{d^d \vec k}{(2 \pi)^{d}}
\tilde{D}_{ij}(\vec k) \exp (i \vec k \cdot \vec x) \label{COR}
\end{eqnarray}
is convenient to be parameterized in the following way \cite{RUB,ADZ2}:
\begin{eqnarray}
\tilde{D}_{ij}(\vec k)=g_0 \nu_0^3 k^{4-d-2\epsilon}
[(1+\alpha_{10} \xi_k^2) P_{ij}(\vec k) + \alpha_{20} R_{ij}(\vec k)]\,,
\label{D}
\end{eqnarray}
where a vector $\vec k$ is the wave vector, $d$ is the dimension of
the space (in our case: $2 \leq d$), $\epsilon \geq 0$ is a
dimensionless parameter of the model. If the dimension of the system
is taken $d>2$, then the physical value of this parameter is
$\epsilon =2$ (the so-called energy pumping regime). The situation is
more complicated when $d=2$. In this case the new integrals of
motion arise, namely the enstrophy, and all its powers (for
details see Ref.\,\cite{FOR}) which leads to ambiguity in
determination of the inertial range and this freedom is given in the RG
method by the value of the parameter $\epsilon$. The value
$\epsilon =3$ corresponds to the so-called enstrophy pumping regime.
This problem of uncertainty cannot be solved in the
framework of the RG technique. On the other hand, the value of
$\epsilon$ is not important for stability of the fixed point when
$d=2$. Thus, it is not important, from our point of view, what is
the value of $\epsilon$ in the case $d=2$. Its  value $\epsilon
=0$ corresponds to a logarithmic perturbation theory for
calculation of Green functions when $g_0$, which plays the role
of a bare coupling constant of the model, becomes
dimensionless. The problem of continuation from $\epsilon=0$
to physical values has been discussed in \cite{ADZ3}.
The $(d \times d)$-matrices $P_{ij}$ and $R_{ij}$ are the
transverse projection operators and in the wave-number space are
defined by the relations
\begin{eqnarray}
P_{ij}(\vec k)= \delta_{ij}-\frac{k_i k_j}{k^2}\,,\,\,\,
R_{ij}(\vec k)= \left(n_i-\xi_k \frac{k_i}{k}\right)
\left(n_j-\xi_k \frac{k_j}{k}\right)\,,
\label{pro}
\end{eqnarray}
where $\xi_k$ is given by the equation
$\xi_k=\vec k \cdot \vec n /k\,.$
In eq.(\ref{pro}) the unit vector $\vec n$ specifies the direction
of the anisotropy axis. The tensor $\tilde{D}_{ij}$ given by
eq.(\ref{D}) is the most general form with respect to the
condition of incompressibility of the system under consideration
and contains two dimensionless free parameters $\alpha_{10}$ and
$\alpha_{20}$. From the positiveness of the correlator tensor
$D_{ij}$ one immediately gets restrictions on the above
parameters, namely $\alpha_{10} \geq -1$ and $\alpha_{20} \geq 0$.
In what follows we assume that these parameters are small enough
and generate only small deviations from the isotropy case.

Using the well-known Martin--Siggia-Rose formalism
of the stochastic quantization
\cite{Domin,MSR,Pismak,Janssen}
one can transform the stochastic problem (\ref{NS})
with the correlator (\ref{COR})
into the quantum field model of the fields
$\vec v$ and $\vec{v}^{,}$.
Here $\vec{v}^{,}$ is an independent of the $\vec v$ auxiliary
incompressible field which we have to introduce when transforming the
stochastic problem into the functional form.

The action of the fields $\vec v$ and $\vec{v}^{,}$
is given in the form
\begin{eqnarray}
S&=&\frac{1}{2} \int d^d \vec x_1 d t_1 d^d \vec x_2 d t_2
\left[v^{,}_{i}(\vec x_1 ,t_1) D_{ij}(\vec x_1 -\vec x_2 , t_1-t_2)
v^{,}_{j}(\vec x_2 ,t_2)\right] \nonumber \\
&+& \int d^d \vec x d t \left\{\vec{v}^{,}(\vec x ,t)
\left[-\partial_t \vec v -(\vec v \cdot \vec \nabla)\vec v +
\nu_0 \vec{\nabla}^2 \vec v +\vec f^A \right](\vec x ,t)  \right\}\,.
\label{action1}
\end{eqnarray}
The functional formulation gives the possibility of using the quantum
field theory methods including the RG technique to solve the problem.
By means of the RG approach it is possible to extract large-scale
asymptotic behaviour of the correlation functions after an appropriate
renormalization procedure which is needed to remove UV-divergences.

Now we can return back to give an explicit form of the
anisotropic dissipative term $\vec f^{A}$. When $d>2$
the UV-divergences  are only present  in the
one-particle-irreducible Green function $<\vec v^{,} \vec v >$.
To remove them, one
needs to introduce into the action in addition to the counterterm
$\vec v^{,} \vec{\nabla}^2 \vec v$ (the only counterterm needed in
the isotropic model) the following ones $\vec v^{,} (\vec n \cdot
\nabla)^2 \vec v$, $(\vec n \cdot \vec v^{,})
\vec{\nabla}^2 (\vec n \cdot \vec v)$ and
$(\vec n \cdot \vec v^{,}) (\vec n \cdot \vec{\nabla})^2
(\vec n \cdot \vec
v)$. These additional terms are needed to remove divergences
related to anisotropic structures. In this case ($d>2$), one can use
the above action (\ref{action1}) with
(\ref{D}) to solve the anisotropic turbulent problem.
Therefore, in order to arrive at the multiplicative renormalizable
model, we have to take the term $\vec f^A$
in the form
\begin{eqnarray}
\vec f^A=\nu_0 \left[\chi_{10}(\vec n \cdot \vec \nabla)^2 \vec v +
\chi_{20} \vec n \vec{\nabla}^2 (\vec n \cdot \vec v) +
\chi_{30} \vec n (\vec n \cdot \vec \nabla)^2 (\vec n \cdot
\vec v) \right]\,.
\label{sila}
\end{eqnarray}
Bare parameters $\chi_{10}$, $\chi_{20}$ and $\chi_{30}$
characterize the weight of the individual structures in
(\ref{sila}).

A more complicated situation arises in the specific  case $d=2$
where new divergences appear. They are related to the
1-irreducible Green function $<\vec v^{,} \vec v^{,}>$ which is
finite when $d>2$. Here one comes to a problem how to remove these
divergences because the term in our action, which contains
a structure of this type is nonlocal, namely $\vec v^{,} k^{4-d-2
\epsilon} \vec v^{,}$. The only correct way of solving the above
problem is to introduce into the action a new local term of the form
$\vec v^{,} \vec{\nabla}^2 \vec v^{,}$ (isotropic case)
\cite{HON}. In the anisotropic case, we have to introduce
additional counterterms $\vec v^{,} (\vec n \cdot \nabla)^2 \vec
v^{,}$, $(\vec n \cdot \vec v^{,}) \vec{\nabla}^2 (\vec n \cdot
\vec v^{,})$ and $(\vec n \cdot \vec v^{,}) (\vec n \cdot
\vec{\nabla})^2 (\vec n \cdot \vec v^{,})$. In
\cite{HON,RON}  a double-expansion method with a simultaneous
deviation $2\delta=d-2$ from the spatial dimension $d=2$ and also
a deviation $\epsilon$ from the $k^2$ form of the forcing pair
correlation function proportional to $k^{2-2 \delta-2 \epsilon}$
was proposed. We shall follow the formulation founded on the
two-expansion parameters in the present paper.

In this case, the kernel (\ref{D}) corresponding to the correlation matrix
$D_{ij}(\vec x_1 - \vec x_2, t_2-t_1)$ in action (\ref{action1}) is
replaced by the expression
\begin{eqnarray}
\tilde{{ D}}_{ij}(\vec k)&=&g_{10} \nu_0^3 k^{2-2\delta
-2\epsilon} [(1+\alpha_{10} \xi_k^2) P_{ij}(\vec k) + \alpha_{20}
R_{ij}(\vec k)] \nonumber \\ &+& g_{20} \nu_0^3 k^2
[(1+\alpha_{30} \xi_k^2) P_{ij}(\vec k)+ (\alpha_{40} +
\alpha_{50} \xi_k^2) R_{ij}(\vec k)] \,.\label{D1}
\end{eqnarray}
Here $P_{ij}$ and $R_{ij}$ are given by relations (\ref{pro}), $g_{20}$,
$\alpha_{30}$, $\alpha_{40}$ and $\alpha_{50}$ are new parameters of
the model, and the parameter $g_0$ in eq.\,(\ref{D}) is now renamed as
$g_{10}$. One can see that in such a formulation
the counterterm $\vec v^{,} \vec{\nabla}^2 \vec v^{,}$
and all anisotropic terms can be taken into
account by renormalization of the coupling constant $g_{20}$,
and the parameters $\alpha_{30}$, $\alpha_{40}$ and $\alpha_{50}$.

The action (\ref{action1}) with the kernel $\tilde{D}_{ij}(\vec k)$
(\ref{D1})
is given in the form convenient for
realization of the quantum field perturbation analysis with
the standard Feynman diagram technique. From the quadratic  part of the
action one obtains the matrix of bare propagators
(in the wave-number -
frequency representation) \\ \vspace{5mm} \unitlength=1.00mm
\special{em:linewidth 0.4pt} \linethickness{0.4pt}
\begin{picture}(96.33,29.94)
\put(27.33,21.93){\line(-1,0){19.33}}
\put(30.67,21.93){\makebox(0,0)[ lc]
{$ =<{ v_i v_j}>_0\equiv \Delta_{ij}^{{ vv}}(\vec k , \omega_k), $}}
\put(8.00,12.87){\line(1,0){19.33}}
\put(25.00,14.30){\line(0,-1){3.00}}
\put(30.67,12.87){\makebox(0,0)[ lc]
{$ =<{ v_i v^{,}_j}>_0\equiv
\Delta_{ij}^{{ v v^{,}}}(\vec k , \omega_k), $}}
\put(8.00,3.87){\line(1,0){19.33}}
\put(25.00,5.30){\line(0,-1){3.00}}
\put(11.00,5.30){\line(0,-1){3.00}}
\put(30.67,3.87){\makebox(0,0)[ lc]
{$ =<{ v^{,}_i v^{,}_j}>_0\equiv
\Delta_{ij}^{{ v^{,} v^{,}}}(\vec k , \omega_k)=0, $}}
\end{picture}\\
\vspace{5mm}
where
\begin{eqnarray}
\Delta_{ij}^{vv}(\vec k , \omega_k)&=& -\frac{K_3}{K_1 K_2} P_{ij}
\nonumber \\
&+& \frac{1}{K_1 (K_2+\tilde K (1-\xi_k^2))} \bigg[
\frac{\tilde K K_3}{K_2}+
\frac{\tilde K (K_3 + K_4 (1-\xi_k^2))}{(K_1+\tilde K (1-\xi_k^2))}-K_4
\bigg] R_{ij} \nonumber \\
\Delta_{ij}^{v v^{,}}(\vec k , \omega_k)&=& \frac{1}{K_2} P_{ij} -
\frac{\tilde K}{K_2 (K_2+\tilde K (1-\xi_k^2))} R_{ij}\,,
\end{eqnarray}
with
\begin{eqnarray}
K_1&=& i \omega_k + \nu_{0}k^2 + \nu_0 \chi_{10} (\vec n \cdot
\vec k)^2\,, \nonumber \\ K_2&=& - i \omega_k + \nu_{0}k^2 + \nu_0
\chi_{10} (\vec n \cdot \vec k)^2\,, \nonumber \\ K_3&=&
-g_{10}\nu_0^3 k^{2-2\delta -2\epsilon} (1+\alpha_{10}\xi_k^2) -
g_{20}\nu_0^3 k^2 (1+\alpha_{30}\xi_k^2)\,, \nonumber \\ K_4&=&
-g_{10}\nu_0^3 k^{2-2\delta -2\epsilon} \alpha_{20} - g_{20}\nu_0^3
k^2 (\alpha_{40}+\alpha_{50}\xi_k^2)\,, \nonumber \\ \tilde K &=&
\nu_0 \chi_{20} k^2 + \nu_0 \chi_{30} (\vec n \cdot \vec k)^2\,.
\label{parts}
\end{eqnarray}
The propagators are written in the form suitable also for strong
anisotropy when the parameters $\alpha_{i0}$ are not small.
In the case of weak anisotropy, it is possible to make the expansion and
work only with linear terms with respect to all parameters which
characterize anisotropy. The interaction vertex in our model is given
by the expression \vspace{5mm}

\unitlength=1.00mm
\special{em:linewidth 0.4pt}
\linethickness{0.4pt}
\begin{picture}(96.33,20.94)
\put(1.00,12.94){\line(1,0){11.00}}
\put(20.00,5.19){\line(-1,1){8.00}}
\put(12.00,12.94){\line(1,1){8.00}}
\put(8.67,14.66){\line(0,-1){3.87}} \put(4.33,15.09){\makebox(0,0)[
cb]  {$ i $}} \put(14.00,19.82){\makebox(0,0)[ cb]  {$ j $}}
\put(14.00,6.91){\makebox(0,0)[ ct]  {$ l $}}
\put(25.00,12.94){\makebox(0,0)[ lc] {$\equiv V_{ijl}=i( k_j
\delta_{il}+k_l \delta_{ij}).$}}
\end{picture}
\vspace{5mm}

Here, the wave vector $\vec k$ corresponds to the field $\vec v^{,}$.
Now one can use the above introduced Feynman rules for computation of
all needed graphs.

\section{RG-analysis and  Stability of the Fixed Point}

Using the standard analysis of quantum field theory (see e.g.
\cite{ADZ1,VASB,COL,ZIN}), one can find out that the UV divergences
of one-particle-irreducible Green functions $<v v^{,}>_{IR}$ and
$<v^{,} v^{,}>_{IR}$
are quadratic in the wave vector.
The last one takes place only in the case when
dimension of the space is two. All terms needed for removing
the divergences are included in the action (\ref{action1}) with
(\ref{sila}) and kernel (\ref{D1}). This leads to the fact that
our model is multiplicatively renormalizable. Thus, one can
immediately write down the renormalized action in wave-number -
frequency representation with $\vec{\nabla} \rightarrow i\vec k,
\partial_t \rightarrow -i\omega_k$ (all needed integrations and
summations are assumed)
\begin{eqnarray}
S^{R}(v, v^{,})&=& \frac{1}{2} v_i^{,}\Bigg[ g_1 \nu^3 \mu^{2 \epsilon}
k^{2-2\delta -2\epsilon}\left( \left(1+\alpha_1 \xi_k^2\right)P_{ij} +
\alpha_2 R_{ij} \right) \nonumber \\ &+& g_2 \nu^3 \mu^{-2\delta} k^2
\left(\left( Z_5 + Z_6 \alpha_3 \xi_k^2\right) P_{ij} +\left(Z_7 \alpha_4
+Z_8 \alpha_5 \xi_k^2 \right) R_{ij}\right) \Bigg] v_{j}^{,} \nonumber \\
&+& v_i^{,} \Bigg[(i \omega_k -Z_1 \nu k^2)P_{ij}
-\nu k^2 \left(Z_2 \chi_1 \xi_k^2 P_{ij} +
\left(Z_3 \chi_2 + Z_4 \chi_3 \xi_k^2 \right)R_{ij}\right) \Bigg]v_{j}
\nonumber \\
&+& \frac{1}{2}v_i^{,} v_j v_l V_{ijl}\,, \label{action3}
\end{eqnarray}
where $\mu$ is a scale setting parameter with the same canonical
dimension as the wave number. Quantities $g_i$, $\chi_i$,
$\alpha_3$, $\alpha_4$, $\alpha_5$ and $\nu$ are the renormalized
counterparts of bare ones and $Z_i$ are renormalization constants
which are expressed via the UV divergent parts of the functions
$<v v^{,}>_{IR}$ and $<v^{,} v^{,}>_{IR}$. Their general form in one
loop approximation is
\begin{eqnarray}
Z_i=1-F_i\,  \mbox{Poles}_i^{\delta , \epsilon}\,.
\label{zet}
\end{eqnarray}

In the standard MS scheme the amplitudes $F_i$ are only some functions of
 $g_i$, $\chi_i$, $\alpha_3$, $\alpha_4,$ $\alpha_5$  and are independent
of $d$ and $ \epsilon.$ The terms  $\mbox{Poles}_i^{\delta ,
\epsilon}$ are given by linear combinations of the poles
$\frac{1}{\epsilon}$, $\frac{1}{\delta}$ and $\frac{1}{2\epsilon
+\delta}$ (for $\delta \rightarrow 0, \epsilon \rightarrow 0 $).
The amplitudes $F_i=F_i^1 F_i^2$ are a product of two multipliers
$F_i^1, F^2_i.$ One of them, say, $F_i^1$ is a multiplier
originating from the divergent part of the Feynman diagrams, and
the second one $F_i^2$ is connected only with the tensor nature of
the diagrams. We explain that using the following simple example.
Consider an UV-divergent integral $$ I({\bf k}, {\bf n}) \equiv
n_i n_j k_l k_m\int d^d {\bf q} \frac{1}{(q^2+m^2)^{1+2\delta}}
(\frac{q_iq_jq_lq_m}{q^4}
-\frac{\delta_{ij}q_lq_m+\delta_{il}q_jq_m+\delta_{jl}q_iq_m}{3q^2})$$
(summations over repeated indices are implied) where $m$ is an
infrared mass. It can be simplified in the following way: $$
I({\bf k}, {\bf n}) \equiv n_i n_j k_l k_m S_{ijlm}\int_0^{\infty}
d q^2 \frac{q^{2\delta}}{2(q^2+m^2)^{1+2\delta}},$$ where
$$S_{ijlm}=\frac{S_d}{d(d+2)}(\delta_{ij}\delta_{lm}+\delta_{il}\delta_{jm}+
\delta_{im}\delta_{jl}-\frac{(d+2)}{3}(\delta_{ij}\delta_{lm}+
\delta_{il}\delta_{jm}+\delta_{im}\delta_{jl})),$$
$$\int_0^{\infty} d q^2
\frac{q^{2\delta}}{2(q^2+m^2)^{1+2\delta}}=\frac{\Gamma{(\delta
+1)}\Gamma{(\delta)}}{2 m^{2\delta}\Gamma(2\delta+1)}, $$ and
$S_d=2\pi^{d/2}/\Gamma(d/2)$ is the surface of unit the
d-dimensional sphere. The purely UV divergent part manifests
itself as the pole in $2\delta=d-2$; therefore, we find $$\mbox{UV
div. part of}\,\,\,\, I= \frac{1}{2\delta}(F^2_1 k^2+F^2_2({\bf n
k})^2),$$ where $F^2_1=F^2_2/2 = (1-d)S_d/3d(d+2).$

In the standard MS scheme one puts $d=2$ in $F^2_1,F^2_2$; therefore
the d-dependence of these multipliers is ignored. For the theories
with vector fields and, consequently, with tensor diagrams, where
the sign of values of fixed points and/or their stability depend
on the dimension $d$, the procedure, which eliminates the
dependence of multipliers of the type $F^2_1,F^2_2$ on $d,$ is not
completely correct because one is not able to control the
stability of the fixed point when drives to $d=3.$ In the analysis
of Feynman diagrams we propose to slightly modify the MS scheme in
such a way that we keep the d-dependence of $F$ in (\ref{zet}).
The following calculations of RG functions ($\beta-$ functions and
anomalous dimensions) allow one to arrive at the results which are in
qualitative agreement with the results obtained recently in
the framework of the simple analytical regularization scheme
\cite{BUSA}, i.e. we obtain the fixed point which is not stable for
$d=2,$ but whose stability is restored for a borderline
dimension $2<d_c<3.$

The transition from the action (\ref{action1}) to the
renormalized one (\ref{action3}) is given by the introduction of
the following renormalization constants  $Z$:
\begin{eqnarray}
\nu_0=\nu Z_{\nu}\,,\,\,
g_{10}=g_1 \mu^{2\epsilon}Z_{g_1}\,,\,\,
g_{20}=g_2 \mu^{-2\delta}Z_{g_2}\,,\,\,
\chi_{i0}=\chi_i Z_{\chi_i}\,,\,\,
\alpha_{(i+2)0}=\alpha_{i+2}Z_{\alpha_{i+2}},\label{ren}
\end{eqnarray}
where $i=1,2,3$. By comparison of the corresponding terms in the
action (\ref{action3}) with definitions of the renormalization
constants $Z$ for the parameters (\ref{ren}), one can immediately
write down relations between them. Namely, we have
\begin{eqnarray}
Z_{\nu}&=&Z_1\,,\nonumber \\
Z_{g_1}&=&Z_{1}^{-3}\,,\nonumber \\
Z_{g_2}&=&Z_{5}Z_1^{-3}\,, \nonumber \\
Z_{\chi_{i}}&=&Z_{1+i}Z_{1}^{-1}\,,\nonumber \\
Z_{\alpha_{i+2}}&=&Z_{i+5}Z_{5}^{-1}\,, \label{rel}
\end{eqnarray}
where again $i=1,2,3$.

In the one-loop approximation, divergent one-irreducible Green
functions \\ $<v^{,} v>_{IR}$ and $<v^{,} v^{,}>_{IR}$ are
represented by the Feynman graphs
\begin{eqnarray}
\unitlength=1.00mm
\special{em:linewidth 0.4pt}
\linethickness{0.4pt}
\begin{picture}(96.67,8.74)
\put(5.00,2.72){\makebox(0,0)[ rc] {$<v^{,}v^{,}>_{IR}=$}}
\put(21.50,2.73){\oval(15.00,10.32)[ t]  }
\put(21.50,2.73){\oval(15.00,10.32)[ b]  }
\put(29.00,2.73){\line(1,0){7.00}}
\put(7.00,2.73){\line(1,0){7.00}} \put(38.00,2.72){\makebox(0,0)[
rc]  {,}} \put(63.67,2.72){\makebox(0,0)[ rc]  {$<v^{,}v>_{IR}=$}}
\put(80.17,2.73){\oval(15.00,10.32)[ t]  }
\put(80.17,2.73){\oval(15.00,10.32)[ b]  }
\put(87.67,2.73){\line(1,0){7.00}}
\put(65.67,2.73){\line(1,0){7.00}}
\put(10.67,0.57){\line(0,1){4.30}}
\put(32.33,4.87){\line(0,-1){4.30}}
\put(69.33,0.57){\line(0,1){4.30}}
\put(84.67,5.73){\line(0,1){3.01}}
\end{picture} .
\label{obr1}
\end{eqnarray}
The divergent parts of these diagrams
$\Gamma^{v^{,}v^{,}},$ $ \Gamma^{v^{,}v}$ have the structure
\begin{eqnarray}
\Gamma^{v^{,}v^{,}}&=&\frac{1}{2}\nu^3 A \nonumber \\
&\times&
\Bigg[\frac{g_1^2}{4\epsilon +2\delta} \left(a_1 \delta_{ij} k^2+
a_2 \delta_{ij} (\vec n \cdot \vec k)^2+
a_3 n_i n_j k^2 + a_4 n_i n_j (\vec n \cdot \vec k)^2\right) \nonumber \\
&+&\frac{g_1 g_2}{2\epsilon}
\left(b_1 \delta_{ij} k^2+
b_2 \delta_{ij} (\vec n \cdot \vec k)^2+
b_3 n_i n_j k^2 + b_4 n_i n_j (\vec n \cdot \vec k)^2\right) \nonumber \\
&+&\frac{g_2^2}{-2\delta}
\left(c_1 \delta_{ij} k^2+
c_2 \delta_{ij} (\vec n \cdot \vec k)^2+
c_3 n_i n_j k^2 + c_4 n_i n_j (\vec n \cdot \vec k)^2\right)
\Bigg]\,,\nonumber \\
\Gamma^{v^{,}v}&=&-\nu A \nonumber \\
&\times& \Bigg[
\frac{g_1}{2\epsilon} \left(d_1 \delta_{ij} k^2+
d_2 \delta_{ij} (\vec n \cdot \vec k)^2+
d_3 n_i n_j k^2 + d_4 n_i n_j (\vec n \cdot \vec k)^2\right) \nonumber \\
&+&\frac{g_2}{-2\delta} \left(e_1 \delta_{ij} k^2+
e_2 \delta_{ij} (\vec n \cdot \vec k)^2+
e_3 n_i n_j k^2 + e_4 n_i n_j (\vec n \cdot \vec k)^2\right) \Bigg]\,,
\end{eqnarray}
where parameter $A$ and functions $a_i$, $b_i$, $c_i$, $d_i$ and
$e_i$ are given in the Appendix I ($i=1...4$).
The counterterms are built up from these divergent
parts  which lead to the following equations for
renormalization constants:
\begin{eqnarray}
Z_1&=& 1- A \left( \frac{g_1}{2\epsilon} d_1 - \frac{g_2}{2\delta}
e_1\right)\,, \nonumber \\ Z_{1+i}&=& 1- \frac{A}{\chi_i}
\left(\frac{g_1}{2\epsilon} d_{1+i} - \frac{g_2}{2\delta}
e_{1+i}\right)\,, \nonumber \\ Z_5&=& 1+
\frac{A}{2}\left(\frac{g_1^2}{g_2}\frac{a_1}{4\epsilon +2\delta} +
\frac{g_1}{2\epsilon} b_1 - \frac{g_2}{2\delta} c_1\right)\,,
\nonumber \\ Z_{5+i} &=& 1+ \frac{A}{2 \alpha_{i+2}} \left(
\frac{g_1^2}{g_2}\frac{a_{i+1}}{4\epsilon +2\delta} +
\frac{g_1}{2\epsilon} b_{i+1} - \frac{g_2}{2\delta}
c_{i+1}\right)\,,\nonumber \\ i&=&1,2,3\,.
\end{eqnarray}

From these expressions one can define the
corresponding anomalous dimensions
$\gamma_i = \mu \partial_{\mu} \ln Z_{i}$
for all renormalization constants $Z_i$ (logarithmic derivative
$\mu \partial_{\mu}$ is taken at fixed values of  all bare
parameters). The $\beta$-functions  for all invariant charges
(running coupling constants $g_1$, $g_2$, and parameters $\chi_i$,
$\alpha_{i+2}$) where $i=1,2,3$ are given by the following
relations: $\beta_{g_i}=\mu \partial_{\mu}g_i$ ($i=1,2$),
$\beta_{\chi_i}=\mu
\partial_{\mu} \chi_i$ and $\beta_{\alpha_{i+2}}=\mu
\partial_{\mu} \alpha_{i+2}$ ($i=1,2,3$). Now using equations
(\ref{rel})  and definitions  given above one can immediately
write the $\beta$-functions in the form
\begin{eqnarray}
\beta_{g_1}&=&-g_1 (2\epsilon +\gamma_{g_1})
            =g_1 (-2\epsilon + 3 \gamma_1)\,, \nonumber \\
\beta_{g_2}&=&\,\,\,\,\,g_2 (2\delta - \gamma_{g_2})
            =g_2 (2\delta +3
\gamma_{1}-\gamma_5)\,, \nonumber \\
\beta_{\chi_i}&=&-\chi_i \gamma_{\chi_i}
        \qquad\,\,\,\,\,\, =
             -\chi_i (\gamma_{i+1}-\gamma_1)\,, \nonumber \\
\beta_{\alpha_{i+2}}&=&-\alpha_{i+2} \gamma_{\alpha_{i+2}}
              \,\,\,\,\,\,\,\, =
-\alpha_{i+2} (\gamma_{i+5}-\gamma_5)\,,\,\,\,\, i=1,2,3\,,
\end{eqnarray}
where
\begin{eqnarray}
\gamma_1&=& A (g_1 d_1 + g_2 e_1)\,,
\nonumber \\
\gamma_{i+1}&=& \frac{A}{\chi_i} \left(g_1 d_{i+1} + g_2 e_{i+1}\right)\,,
\nonumber \\
\gamma_5&=& -\frac{A}{2}\left(\frac{g_1^2}{g_2} a_1 + g_1 b_1 + g_2
c_1\right)\,,
\nonumber \\
\gamma_{i+5}&=&-\frac{A}{2 \alpha_{i+2}}\left(\frac{g_1^2}{g_2} a_{i+1}+
g_1 b_{i+1} + g_2 c_{i+1}\right)\,,\,\,\,i=1,2,3\,.
\label{gamma}
\end{eqnarray}
By substitution of the anomalous dimensions $\gamma_i$
(\ref{gamma}) into the expressions for the $\beta$-functions one
obtains
\begin{eqnarray}
\beta_{g_1}&=&g_1 (-2\epsilon + 3 A (g_1 d_1 + g_2 e_1))\,, \nonumber \\
\beta_{g_2}&=&g_2 \left[2\delta + 3 A (g_1 d_1 + g_2 e_1)+
 \frac{A}{2}\left(\frac{g_1^2}{g_2} a_1 +
g_1 b_1 + g_2 c_1\right) \right]\,,
\nonumber \\
\beta_{\chi_i}&=&- A \left[\left(g_1 d_{i+1} + g_2 e_{i+1}\right)-
 \chi_i (g_1 d_1 + g_2 e_1)\right]\,, \nonumber \\
\beta_{\alpha_{i+2}}&=&
-\frac{A}{2} \left[-\left(\frac{g_1^2}{g_2} a_{i+1}+
g_1 b_{i+1} + g_2 c_{i+1}\right) +
\alpha_{i+2}\left(\frac{g_1^2}{g_2} a_1 + g_1 b_1 + g_2 c_1\right)
\right]\,,
\nonumber \\
i&=&1,2,3\,.
\end{eqnarray}

\input epsf
   \begin{figure}[t]
     \vspace{-0.5cm}
       \begin{center}
       \leavevmode
       \epsfxsize=10cm
       \epsffile{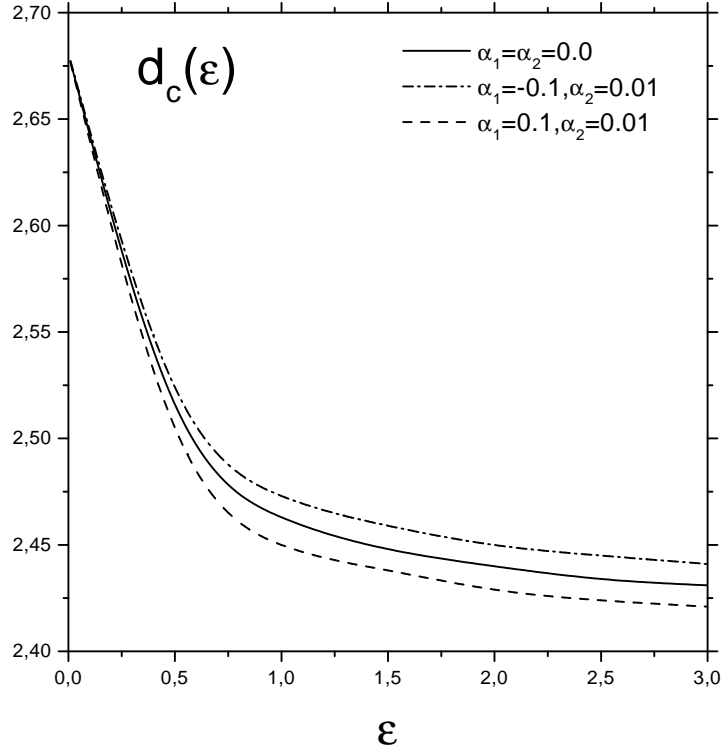}
   \end{center}
\vspace{-1.5cm} \caption{ Dependence of the borderline dimension
$d_c$ on the parameter $\epsilon$ and for concrete values of
$\alpha_1$ and $\alpha_2$. \label{fig1}}
\end{figure}

The fixed point of the RG-equations is defined by the system of eight
equations
\begin{eqnarray}
\beta_{C}(C_{*})=0\,,
\end{eqnarray}
where we denote $C=\{g_1,g_2,\chi_i,\alpha_{i+2}\}, i=1,2,3$ and
$C_{*}$ is the corresponding  value for the fixed point.
The IR stability of the fixed point is determined by the positive
real parts of the eigenvalues of the matrix
\begin{eqnarray}
\omega_{lm}=\left(\frac{\partial \beta_{C_l}}{\partial
C_m}\right)_{C=C_{*}}\, \,\, l,\, m =1,...8. \label{mat}
\end{eqnarray}

\input epsf
   \begin{figure}[t]
     \vspace{-0.5cm}
       \begin{center}
       \leavevmode
       \epsfxsize=10cm
       \epsffile{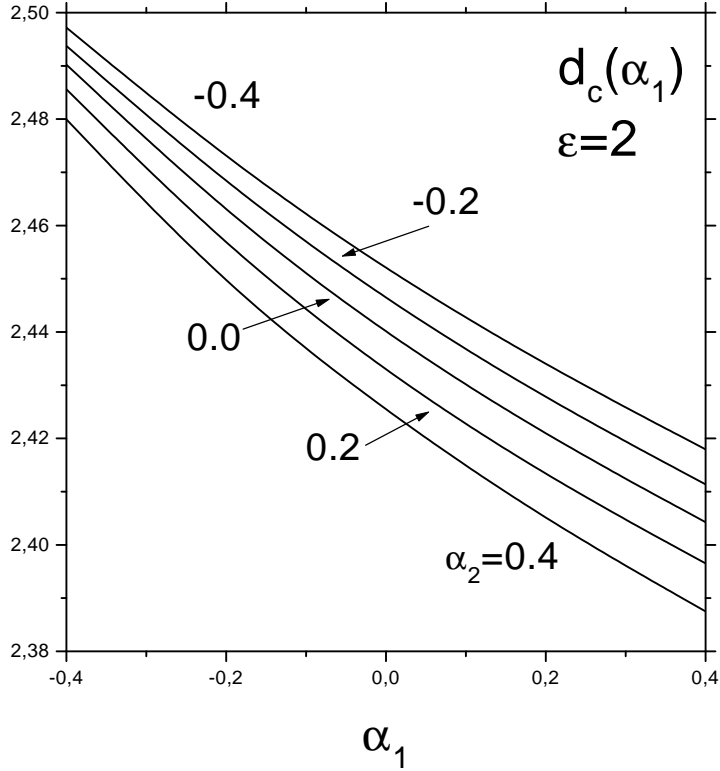}
   \end{center}
\vspace{-1.5cm} \caption{Dependence of the borderline dimension
$d_c$ on the parameters $\alpha_1$ and $\alpha_2$ for physical
value $\epsilon=2$. \label{fig2}}
\end{figure}
Now we have all necessary tools at hand to investigate the fixed
points and their stability. In {\it the isotropic case} all
parameters which are connected with the anisotropy are equal to zero,
and one can immediately find the Kolmogorov fixed point, namely:
\begin{eqnarray}
g_{1*}&=&\frac{1}{A}\,\,
\frac{8 (2 + d) \epsilon (2 \epsilon - 3 d (\delta + \epsilon) +
d^2 (3 \delta + 2 \epsilon))}{9(-1 + d)^3 d (1 + d)(\delta +
\epsilon)}\,,\nonumber \\
g_{2*}&=&\frac{1}{A}\,\,
\frac{8 (-4-2 d+ 2 d^2 +d^3)\epsilon^2}{9(-1 + d)^3 d (1 + d)(\delta +
\epsilon)}\,, \label{IZO}
\end{eqnarray}
where $\delta=(d-2)/2$ and the corresponding $\omega_{ij}$ matrix
has the following eigenvalues:
\begin{eqnarray}
\lambda_{1,2}&=&\frac{1}{6 d (d-1)}\Bigg\{6 d \delta (d-1)+4\epsilon
(2- 3 d + 2 d^2)
\nonumber \\
&\pm& \Bigg[ (6 d \delta (1-d) -4\epsilon (2-3 d + 2 d^2))^2 \nonumber \\
&-& 12 d (d-1) \epsilon (12 d \delta (d-1) +4 \epsilon (2-3 d + 2 d^2))
\Bigg]^{\frac{1}{2}}
\Bigg\}\,.
\end{eqnarray}
By a detail analysis of these eigenvalues we know that in the
interesting region of parameters, namely $\epsilon>0$ and $\delta
\geq0$ (it corresponds to $d \geq 2$) the above computed fixed
point is stable. In the limit $d=2$, this fixed point is in
agreement with that given in \cite{HON,ANT}.

When one considers {\it the weak anisotropy case} the situation
becomes more complicated because of necessity to use all system of
$\beta$-functions if one wants to analyze the stability of the
fixed point. It is also possible to find analytical expressions
for the fixed point in this more complicated case because in the weak
anisotropy limit it is enough to calculate linear corrections of
$\alpha_1$ and $\alpha_2$ to all the quantities (see Appendix II).

To investigate the stability of the fixed point it is necessary to
apply it in the  matrix (\ref{mat}). Analysis of this matrix shows
us that it can be written in the block-diagonal form: $(6\times
6)(2\times 2)$. The $(2\times 2)$ part is given by the
$\beta$-functions of the parameters $\alpha_5$ and $\chi_3$ and,
namely, this block is responsible for the existence of the borderline
dimension $d_c$ because one of its eigenvalues, say
$\lambda_1(\epsilon,d,\alpha_1,\alpha_2)$, has a solution $d_c \in
\langle 2, 3\rangle$ of the equation
$\lambda_1(\epsilon,d_c,\alpha_1,\alpha_2)=0$  for the defined values
of $\epsilon, \alpha_1, \alpha_2$. The following procedure has
been used to find the fixed point:
First we have used the isotropic solution to $g_1$ and $g_2$ to compute
the expressions for $\alpha_{i+2}$ and $\chi_i$, $i=1,2,3$. From
equations $\beta_{\alpha_{5}}=0$ and $\beta_{\chi_3}=0$ one can
immediately find that $\alpha_{5*}=0$ and $\chi_{3*}=0$. After
this we can calculate expressions for the fixed point of the
parameters $\alpha_{i+2}$ and $\chi_i$, $i=1,2$. At the end, we
come back to the equations for $g_1$ and $g_2$, namely $\beta_{g_1}=0$
and $\beta_{g_2}=0$, and find linear corrections of $\alpha_1$ and
$\alpha_2$ to the fixed point. The corresponding expressions for the
fixed point and the corresponding eigenvalue of the stability matrix
responsible for instability are given in Appendix II.

From a numerical analysis of the stability matrix one can find that
in some region of space dimensions $d$ the stability is lost by the
influence of the weak anisotropy. On the other hand, the borderline
dimension $d_c$ arises when going from dimension $d=2$ to $d=3$.
For the energy pumping regime ($\epsilon=2$) we found the critical
dimension $d_c=2.44$. This value corresponds to
$\alpha_1=\alpha_2=0$. This is the case when one supposes only the
fact of anisotropy. Using nonzero values of $\alpha_1$ and
$\alpha_2$ one can also estimate the influence of these parameters
on the borderline dimension $d_c$. It is interesting to calculate
the dependence of  $d_c$ on the parameter $\epsilon$ too. In
Fig.\,\ref{fig1}, this dependence and the dependence on small values of
$\alpha_1$ and $\alpha_2$ are presented. As one can see from this
figure $d_c$ increases when $\epsilon \rightarrow 0$ and also
the parameters $\alpha_1$ and $\alpha_2$ give small corrections to
$d_c$. In Fig.\,\ref{fig2}, the dependence of $d_c$ on $\alpha_1$
and $\alpha_2$ for $\epsilon=2$ is presented.

\section{Conclusion}

We have investigated the influence of the weak anisotropy on the
fully developed turbulence using the quantum field RG double expansion
method and introduced the modified minimal substraction scheme
in which  the space dimension dependence of the divergent parts
of the Feynman
diagrams is kept . We affirm that such a modified approach is
correct when one needs to compute the $d-$dependence of the important
quantities and is necessary for restoration of the stability of
scaling regimes when one makes transition from dimension $d=2$ to
$d=3$. We have derived analytical expressions for the fixed point
in the limit of the weak anisotropy and found the equation which
manages the stability of this point as a function of the parameters
$\epsilon, \alpha_1$ and $\alpha_2$, and allows one to calculate
the borderline dimension $d_c$. Below this dimension the fixed
point is unstable. In the limit case of infinitesimally small
anisotropy ($\alpha_1\rightarrow 0$ and $\alpha_2\rightarrow 0$)
and in the energy pumping regime ($\epsilon =2$) we have found the
borderline dimension $d_c=2.44$. We have also investigated  the
$\epsilon$-dependence of $d_c$ for different values of the
anisotropy parameters $\alpha_1, \alpha_2$ and also the dependence of
$d_c$ on the relatively small values of $\alpha_1$ and $\alpha_2$
for the physical value $\epsilon=2$.

\section*{Acknowledgements}

This work was partially supported by Slovak Academy of Sciences
within the project 7232.

\section*{Appendix I}
The explicit form of the parameter $A$ and functions $a_i$, $b_i$,
$c_i$, $d_i$ and $e_i$ ($i=1...4$) for the divergent parts of
diagrams (\ref{obr1})
\begin{eqnarray}
a_1 &=& \frac{1}{2 d (2 + d) (4 + d) (6 + d)} \nonumber \\
 &\times&
[-48 - 20 d + 70 d^2 + 30 d^3 - 21 d^4 - 10 d^5 - d^6 \nonumber \\
&+& \alpha_2 (24 + 16 d - 22 d^2 - 16 d^3 - 2 d^4)
+ \alpha_1 (24 + 52 d - 4 d^2 - 50 d^3 - 20 d^4 - 2 d^5) \nonumber \\
&+& \chi_1 (-36 - 78 d + 6 d^2 + 75 d^3 + 30 d^4 + 3 d^5) \nonumber \\
&+& \chi_2 (-36 - 24 d + 33 d^2 + 24 d^3 + 3 d^4)
+ \chi_3 (-36 - 9 d + 36 d^2 + 9 d^3) ]\,, \nonumber \\
a_2 &=& \frac{1}{4 d (2 + d) (4 + d) (6 + d)} \nonumber \\
&\times&
[\alpha_1 (-96 - 64 d + 88 d^2 + 64 d^3 + 8 d^4) \nonumber \\
&+& \alpha_2 (-96 - 64 d + 124 d^2 + 82 d^3 - 26 d^4 - 18 d^5 - 2 d^6)
\nonumber \\
&+& \chi_1 (144 + 96 d - 132 d^2 - 96 d^3 - 12 d^4) \nonumber \\
&+& \chi_2 (144 + 96 d - 186 d^2 - 123 d^3 + 39 d^4 + 27 d^5 + 3 d^6)
\nonumber \\
&+& \chi_3 (72 + 6 d - 87 d^2 - 9 d^3 + 15 d^4 + 3 d^5)]\,, \nonumber \\
a_3 &=& a_2\,, \nonumber \\
a_4 &=& \frac{6 \chi_3 (1 - d^2)}{(2 + d) (6 + d)}\,, \nonumber \\
b_1&=& \frac{1}{d (2 + d) (4 + d) (6 + d)} \nonumber \\
&\times&
[-48 - 20 d + 70 d^2 + 30 d^3 - 21 d^4 - 10 d^5 - d^6 +
    \alpha_5 (12 + 3 d - 12 d^2 - 3 d^3) \nonumber \\
&+& (\alpha_2 + \alpha_4) (12 + 8 d - 11 d^2 - 8 d^3 - d^4)  \nonumber \\
&+& (\alpha_1 + \alpha_3) (12 + 26 d - 2 d^2 - 25 d^3 - 10 d^4 - d^5)
\nonumber \\
&+& \chi_1 (-36 - 78 d + 6 d^2 + 75 d^3 + 30 d^4 + 3 d^5) \nonumber \\
&+& \chi_2 (-36 - 24 d + 33 d^2 + 24 d^3 + 3 d^4) +
    \chi_3 (-36 - 9 d + 36 d^2 + 9 d^3)]\,, \nonumber \\
b_2&=&\frac{1}{2 d (2 + d) (4 + d) (6 + d)} \nonumber \\
&\times&
[(\alpha_1 + \alpha_3) (-48 - 32 d + 44 d^2 + 32 d^3 + 4 d^4) \nonumber \\
&+& \alpha_5 (-24 - 2 d + 29 d^2 + 3 d^3 - 5 d^4 - d^5) \nonumber \\
&+& (\alpha_2 + \alpha_4) (-48 - 32 d + 62 d^2 + 41 d^3 - 13 d^4
- 9 d^5 - d^6)
\nonumber \\
&+& \chi_1 (144 + 96 d - 132 d^2 - 96 d^3 - 12 d^4) \nonumber \\
&+& \chi_2 (144 + 96 d - 186 d^2 - 123 d^3 + 39 d^4 + 27 d^5 + 3 d^6)
\nonumber \\
&+& \chi_3 (72 + 6 d - 87 d^2 - 9 d^3 + 15 d^4 + 3 d^5)]\,,\nonumber \\
b_3 &=& b_2\,, \nonumber \\
b_4 &=& \frac{4 (d^2 -1) (\alpha_5 - 3 \chi_3)}{(2 + d) (6 + d)}\,,
\nonumber \\
c_1 &=& \frac{1}{2 d (2 + d) (4 + d) (6 + d)} \nonumber \\
&\times&
[-48 - 20 d + 70 d^2 + 30 d^3 - 21 d^4 - 10 d^5 - d^6 \nonumber \\
&+& \alpha_5 (24 + 6 d - 24 d^2 - 6 d^3) +
    \alpha_4 (24 + 16 d - 22 d^2 - 16 d^3 - 2 d^4) \nonumber \\
&+& \alpha_3 (24 + 52 d - 4 d^2 - 50 d^3 - 20 d^4 - 2 d^5) \nonumber \\
&+& \chi_1 (-36 - 78 d + 6 d^2 + 75 d^3 + 30 d^4 + 3 d^5) \nonumber \\
&+& \chi_2 (-36 - 24 d + 33 d^2 + 24 d^3 + 3 d^4) +
    \chi_3 (-36 - 9 d + 36 d^2 + 9 d^3)]\,, \nonumber \\
c_2 &=& \frac{1}{4 d (2 + d) (4 + d) (6 + d)} \nonumber \\
&\times&
[\alpha_3 (-96 - 64 d + 88 d^2 + 64 d^3 + 8 d^4) \nonumber \\
&+& \alpha_5 (-48 - 4 d + 58 d^2 + 6 d^3 - 10 d^4 - 2 d^5)\nonumber \\
&+& \alpha_4 (-96 - 64 d + 124 d^2 + 82 d^3 - 26 d^4 - 18 d^5 - 2 d^6)
\nonumber \\
&+& \chi_1 (144 + 96 d - 132 d^2 - 96 d^3 - 12 d^4) \nonumber \\
&+& \chi_2 (144 + 96 d - 186 d^2 - 123 d^3 + 39 d^4 + 27 d^5 + 3 d^6)
\nonumber \\
&+& \chi_3 (72 + 6 d - 87 d^2 - 9 d^3 + 15 d^4 + 3 d^5)]\,, \nonumber \\
c_3 &=& c_2\,, \nonumber \\
c_4 &=& \frac{(d^2 - 1) (4 \alpha_5 - 6 \chi_3)}{(2 + d) (6 + d)}\,,
\nonumber \\
d_1 &=& \frac{1}{4 d (2 + d) (4 + d) (6 + d)} \nonumber \\
&\times&
[24 d - 14 d^2 - 33 d^3 + 13 d^4 + 9 d^5 + d^6 +
    \alpha_2 (12 - 4 d - 13 d^2 + 4 d^3 + d^4) \nonumber \\
&+& \alpha_1 (-12 - 20 d + 3 d^2 + 19 d^3 + 9 d^4 + d^5) \nonumber \\
&+& \chi_1 (36 + 42 d - 18 d^2 - 40 d^3 - 18 d^4 - 2 d^5) \nonumber \\
&+& \chi_2 (-12 + 16 d + 15 d^2 - 16 d^3 - 3 d^4) +
\chi_3 (6 + 9 d - 6 d^2 - 9 d^3)]\,, \nonumber \\
d_2 &=& \frac{1}{8 d (2 + d) (4 + d) (6 + d)} \nonumber \\
&\times&
[\alpha_1 (-48 + 16 d + 52 d^2 - 16 d^3 - 4 d^4) \nonumber \\
&+& \alpha_2 (48 + 80 d - 60 d^2 - 96 d^3 + 10 d^4 + 16 d^5 + 2 d^6)
\nonumber \\
&+& \chi_1 (48 - 64 d - 60 d^2 + 64 d^3 + 12 d^4) \nonumber \\
&+& \chi_2 (-48 - 104 d + 62 d^2 + 127 d^3 - 11 d^4 - 23 d^5 - 3 d^6)
\nonumber \\
&+& \chi_3 (-2 d + 7 d^2 + 5 d^3 - 7 d^4 - 3 d^5)]\,, \nonumber \\
d_3 &=& \frac{1}{8 d (2 + d) (4 + d) (6 + d)} \nonumber \\
&\times&
[\alpha_1 (48 + 56 d - 40 d^2 - 56 d^3 - 8 d^4) +
    \alpha_2 (-48 - 56 d + 40 d^2 + 56 d^3 + 8 d^4) \nonumber \\
&+& \chi_1 (-48 - 104 d + 32 d^2 + 104 d^3 + 16 d^4) \nonumber \\
&+& \chi_2 (48 + 32 d - 38 d^2 - 25 d^3 - 9 d^4 - 7 d^5 - d^6)
\nonumber \\
&+& \chi_3 (22 d - d^2 - 21 d^3 + d^4 - d^5)]\,, \nonumber \\
d_4 &=& \frac{\chi_3 (-10 + d + 10 d^2 - d^3)}{2 (2 + d) (6 + d)}\,,
\nonumber \\
e_1 &=& \frac{1}{4 d (2 + d) (4 + d) (6 + d)} \nonumber \\
&\times&
[24 d - 14 d^2 - 33 d^3 + 13 d^4 + 9 d^5 + d^6 + 3 d \alpha_5 (-1 + d^2)
\nonumber \\
&+& \alpha_4 (12 - 4 d - 13 d^2 + 4 d^3 + d^4) \nonumber \\
&+& \alpha_3 (-12 - 20 d + 3 d^2 + 19 d^3 + 9 d^4 + d^5) \nonumber \\
&+& \chi_1 (36 + 42 d - 18 d^2 - 40 d^3 - 18 d^4 - 2 d^5) \nonumber \\
&+& \chi_2 (-12 + 16 d + 15 d^2 - 16 d^3 - 3 d^4) + \chi_3
(6 + 9 d - 6 d^2 - 9 d^3)]\,, \nonumber \\
e_2 &=& \frac{1}{8 d (2 + d) (4 + d) (6 + d)}\nonumber \\
&\times&
[\alpha_3 (-48 + 16 d + 52 d^2 - 16 d^3 - 4 d^4) +
    \alpha_5 (-8 d^2 - 2 d^3 + 8 d^4 + 2 d^5) \nonumber \\
&+& \alpha_4 (48 + 80 d - 60 d^2 - 96 d^3 + 10 d^4 + 16 d^5 + 2 d^6)
\nonumber \\
&+& \chi_1 (48 - 64 d - 60 d^2 + 64 d^3 + 12 d^4) \nonumber \\
&+& \chi_2 (-48 - 104 d + 62 d^2 + 127 d^3 - 11 d^4 - 23 d^5 - 3 d^6)
\nonumber \\
&+& \chi_3 (-2 d + 7 d^2 + 5 d^3 - 7 d^4 - 3 d^5)]\,, \nonumber \\
e_3 &=& \frac{1}{8 d (2 + d) (4 + d) (6 + d)} \nonumber \\
&\times&
[24 d \alpha_5 (-1 + d^2) + \alpha_3 (48 + 56 d - 40 d^2 - 56 d^3 - 8 d^4)
\nonumber \\
&+& \alpha_4 (-48 - 56 d + 40 d^2 + 56 d^3 + 8 d^4) \nonumber \\
&+& \chi_1 (-48 - 104 d + 32 d^2 + 104 d^3 + 16 d^4) \nonumber \\
&+& \chi_2 (48 + 32 d - 38 d^2 - 25 d^3 - 9 d^4 - 7 d^5 - d^6)
\nonumber \\
&+& \chi_3 (22 d - d^2 - 21 d^3 + d^4 - d^5)]\,, \nonumber \\
e_4 &=& \frac{6 \alpha_5 (1 - d^2) + \chi_3 (-10 + d + 10 d^2 - d^3)}
{2 (2 + d) (6 + d)}\,, \nonumber \\
A &=& \frac{S_d}{(2 \pi)^d (d^2-1)}\,, \nonumber
\end{eqnarray}
where $S_d$ is $d$-dimensional sphere given by the following relation:
\begin{eqnarray}
S_d=\frac{2 \pi^{\frac{d}{2}}}{\Gamma(\frac{d}{2})}\,. \nonumber
\end{eqnarray}

\section*{Appendix II}
We present here the explicit analytical expressions for the fixed
point in the weak ani\-so\-tro\-py limit and also the equation which
governs its stability.

The basic form of the fixed point is
\begin{eqnarray}
g_{1*}&=&g_{10*}+g_{11*} \alpha_1 + g_{12*} \alpha_2\,, \nonumber \\
g_{2*}&=&g_{20*}+g_{21*} \alpha_1 + g_{22*} \alpha_2\,, \nonumber \\
\alpha_{3*}&=&e_{11}\alpha_1 + e_{12} \alpha_2\,, \nonumber \\
\alpha_{4*}&=&e_{21}\alpha_1 + e_{22} \alpha_2\,, \nonumber \\
\chi_{1*}&=&e_{31}\alpha_1 + e_{32} \alpha_2\,, \nonumber \\
\chi_{2*}&=&e_{41}\alpha_1 + e_{42} \alpha_2\,, \nonumber \\
\alpha_{5*}&=&0\,, \nonumber \\
\chi_{3*}&=&0\,, \nonumber
\end{eqnarray}
where $g_{10*}$ and $g_{20*}$ are defined in eq.(\ref{IZO}), and
$g_{11*}, g_{12*}, g_{21*}, g_{22*}$ and $e_{ij}, i=1,2,3,4$, $j=1,2$
are functions only of the dimension $d$ and parameters $\epsilon$ and
$\delta=(d-2)/2$. They have the following form:
\begin{eqnarray}
g_{11*}&=&\frac{g_{11n}}{g_{11d}}\,,
g_{12*}=\frac{g_{12n}}{g_{12d}}\,,
g_{21*}=\frac{g_{21n}}{g_{22d}}\,,
g_{22*}=\frac{g_{22n}}{g_{22d}}\,, \nonumber \\
e_{11}&=&\frac{e_{11n}}{e_{d}}\,, e_{12}=\frac{e_{12n}}{e_{d}}\,,
e_{21}=\frac{e_{21n}}{e_{d}}\,, e_{22}=\frac{e_{22n}}{e_{d}}\,,
\nonumber \\ e_{31}&=&\frac{e_{31n}}{g_s e_{d}}\,,
e_{32}=\frac{e_{32n}}{g_s e_{d}}\,, e_{41}=\frac{e_{41n}}{g_s
e_{d}}\,, e_{42}=\frac{e_{42n}}{g_s e_{d}}\,, \nonumber
\end{eqnarray}
where
\begin{eqnarray}
g_{11n}&=& 3 (d^2-1) g_{10*} (d^6 (g_{10*} + g_{20*})
      ( (5 e_{31}-3) g_{10*} - 3 e_{11} g_{20*} + 5 e_{31} g_{20*})
\nonumber \\
&+& 3 d^5 (g_{10*} + g_{20*}) ((-2 + 3 e_{31} + 2 e_{41}) g_{10*} -
       (2 e_{11} + e_{21} - 3 e_{31} - 2 e_{41}) g_{20*}) \nonumber \\
&-& 8 (g_{10*} + g_{20*}) ((-1 + 3 e_{31} - e_{41}) g_{10*} -
      (e_{11} - e_{21} - 3 e_{31} + e_{41}) g_{20*}) \nonumber \\
&+& d^3 (-((g_{10*} + g_{20*}) ((-4 + 9 e_{31} - 6 e_{41}) g_{10*}
\nonumber \\
&+& (-4 e_{11} + 3 (e_{21} + 3 e_{31} - 2 e_{41})) g_{20*})) \nonumber \\
&+& 8 \delta ((-5 + 10 e_{31} + 3 e_{41}) g_{10*} -
       (5 e_{11} + e_{21} - 10 e_{31} - 3 e_{41}) g_{20*})) \nonumber \\
&+& 2 d (-((g_{10*} + g_{20*}) ((-1 + 6 e_{41}) g_{10*} -
      (e_{11} + 3 e_{21} - 6 e_{41}) g_{20})) \nonumber \\
&+&   16 \delta ((-1 + 3 e_{31} - e_{41}) g_{10*} -
        (e_{11} - e_{21} - 3 e_{31} + e_{41}) g_{20*})) \nonumber \\
&+&  d^2 ((g_{10*} + g_{20*}) ((-15 + 34 e_{31}) g_{10*} +
        (-15 e_{11} + 5 e_{21} + 34 e_{31}) g_{20*}) \nonumber \\
&+& 16 \delta ((-4 + 9 e_{31} + 2 e_{41}) g_{10*} \nonumber \\
&+& (-4 e_{11} + 9 e_{31} + 2 e_{41}) g_{20*})) +
    d^4 (8 \delta (-g_{10*} + 2 e_{31} g_{10*} -
e_{11} g_{20*} + 2 e_{31} g_{20*}) \nonumber \\
&-&  (g_{10*} + g_{20*}) ((-10 + 15 e_{31} + 8 e_{41}) g_{10*}
\nonumber \\
&+& (-10 e_{11} - 3 e{21} + 15 e_{31} + 8 e_{41}) g_{20})))\,,
\nonumber \\
g_{11d}&=& 2 d (4 + d) (-15 d^6 (g_{10*} + g_{20*} )^2 + 6 d^7
(g_{10*}  + g_{20*} )^2
          \nonumber \\
&+& 2 (g_{10*}  + g_{20*} ) (16 \epsilon - 3 (g_{10*}  + g_{20*} )) +
    4 d^4 (\epsilon (g_{10*}  - 2 g_{20*} ) \nonumber \\
&+& 6 (g_{10*}  + g_{20*} )^2 + 3 \delta  (2 g_{10*}  + g_{20*} )) +
    d^5 (3 (g_{10*}  + g_{20*} )^2 + 12 \delta  (2 g_{10*}  + g_{20*} )
\nonumber \\
&-& 4 \epsilon (g_{10*}  + 4 g_{20*} )) -
    4 d^3 (6 (g_{10*}  + g_{20*} )^2 - 3 \epsilon (g_{10*}  + 4 g_{20*} )
\nonumber \\
&+& \delta  (8 \epsilon + 9 (2 g_{10*}  + g_{20*} ))) +
    d (15 (g_{10*}  + g_{20*} )^2 - 8 \epsilon (g_{10*}  + 4 g_{20*} )
\nonumber \\
&+& \delta  (-128 \epsilon + 24 (2 g_{10*}  + g_{20*} ))) -
    d^2 (4 \delta  (32 \epsilon + 6 g_{10*}  + 3 g_{20*} ) \nonumber \\
&+& 3 ((g_{10*}  + g_{20*} )^2 + 4 \epsilon (3 g_{10*}  + 2 g_{20*}
))))\,, \nonumber \\
g_{21n}&=& -((-1 + d^2) (-3 (-4 - d + 4 d^2 + d^3) g_{10*}
       (2 (-2 + d^2) g_{10*} \nonumber \\
&-& (4 - 3 d + d^2) g_{20*}) \times \nonumber \\
&\times& ((6 e_{31} + d^2 (-1 + 2 e_{31}) - 2
(1 + e_{41}) + 3 d (-1 + 2 e_{31} + e_{41}))
          g_{10*} \nonumber \\
&-& ((2 + 3 d + d^2) e_{11} \nonumber \\
&+& (-2 + d) e_{21} - 6 e_{31} - 6 d e_{31} -
            2 d^2 e_{31} + 2 e_{41} - 3 d e_{41}) g_{20*}) \nonumber \\
&+& d (4 + d) ((-4 + 6 e_{31} + d^2 (-2 + 3 e_{31}) + 6 e_{41} +
            d (-8 + 12 e_{31} + 3 e_{41})) g_{10*}^2 \nonumber \\
&+& (2 + d^2 (1 - 2 e_{11}) - 4 e_{11} - 4 e_{21} - 6 e_{31} +
            d (1 - 8 e_{11} - 2 e_{21} + 6 e_{31} - 3 e_{41}) \nonumber \\
&+& 18 e_{41}) g_{10*} g_{20*} +
         ((2 + d + d^2) e_{11} + (-10 + d) e_{21} -
            3 (4 e_{31} + 2 d e_{31} + d^2 e_{31} \nonumber \\
&-& 4 e_{41} + 2 d e_{41})) g_{20*}^2)
       (-8 (2 + d) \epsilon + 3 (-1 + d)^2 (1 + d) (2 g_{10*} +
g_{20*}))))\,,\nonumber \\
g_{21d}&=& 3 (-1 + d)^2 d (4 + 5 d + d^2) (-4 - d + 4 d^2 + d^3) g_{10*}
   (2 (-2 + d^2) g_{10*} \nonumber \\
&-& (4 - 3 d + d^2) g_{20*}) +
  d (4 + d)^2 (-8 (2 + d) \epsilon \nonumber \\
&+& 3 (-1 + d)^2 (1 + d) (2 g_{10*} + g_{20*}))
   (d^2 (8 \delta + 3 g_{10*}) - 4 (g_{10*} + g_{20*}) \nonumber \\
&-& 3 d^3 (g_{10*} + 2 g_{20*}) +
     d^4 (g_{10*} + 4 g_{20*}) + d (16 \delta + 3 g_{10*} + 6 g_{20*}))\,,
\nonumber \\
g_{12n}&=& 3 (-1 + d^2) g_{10*} (3 d^5 (g_{10*} + g_{20*})
     ((-1 + 3 e_{32} + 2 e_{42}) g_{10*} \nonumber \\
&-& (2 e_{12} + e_{22} - 3 e_{32} - 2 e_{42}) g_{20*}) -
    8 (g_{10*} + g_{20*}) ((1 + 3 e_{32} - e_{42}) g_{10*} \nonumber \\
&-& (e_{12} - e_{22} - 3 e_{32} + e_{42}) g_{20*}) \nonumber \\
&+& d^6 (g_{10*} + g_{20*}) (-3 e_{12} g_{20*} +
5 e_{32} (g_{10*} + g_{20*}))\nonumber \\
&+& d^3 (-((g_{10*} + g_{20*}) (3 (1 + 3 e_{32} - 2 e_{42}) g_{10*}
\nonumber \\
&+&  (-4 e_{12} + 3 (e_{22} + 3 e_{32} - 2 e_{42})) g_{20*})) \nonumber \\
&+& 8 \delta ((-1 + 10 e_{32} + 3 e_{42}) g_{10*} -
(5 e_{12} + e_{22} - 10 e_{32} - 3 e_{42}) g_{20*})) \nonumber \\
&+& 2 d (-((g_{10*} + g_{20*}) ((-3 + 6 e_{42}) g_{10*} -
     (e_{12} + 3 e_{22} - 6 e_{42}) g_{20*})) \nonumber \\
&+& 16 \delta ((1 + 3 e_{32} - e_{42}) g_{10*} -
        (e_{12} - e_{22} - 3 e_{32} + e_{42}) g_{20*})) \nonumber \\
&+& d^4 (-((g_{10*} + g_{20*}) ((-3 + 15 e_{32} + 8 e_{42}) g_{10*}
\nonumber \\
&+& (-10 e_{12} - 3 e_{22} + 15 e_{32} + 8 e_{42}) g_{20*})) \nonumber \\
&+& 8 \delta (-(e_{12} g_{20*}) + 2 e_{32} (g_{10*} + g_{20*}))) +
    d^2 ((g_{10*} + g_{20*}) ((5 + 34 e_{32}) g_{10*} \nonumber \\
&+& (-15 e_{12} + 5 e_{22} + 34 e_{32}) g_{20*}) +
       16 \delta (9 e_{32} (g_{10*} + g_{20*}) \nonumber \\
&+& 2 (-2 e_{12} g_{20*} + e_{42} (g_{10*} + g_{20*})))))\,, \nonumber \\
g_{12d}&=&g_{11d}\,, \nonumber \\
g_{22n}&=& -((-1 + d^2) (-3 (-4 - d + 4 d^2 + d^3) g_{10*}
       (2 (-2 + d^2) g_{10*} \nonumber\\
&-& (4 - 3 d + d^2) g_{20*})
((2 + 6 e_{32} + 2 d^2 e_{32} - 2 e_{42} +
         d (-1 + 6 e_{32} + 3 e_{42})) g_{10*} \nonumber \\
&-& ((2 + 3 d + d^2) e_{12} + (-2 + d) e_{22} \nonumber \\
&-& 6 e_{32} - 6 d e_{32} -
            2 d^2 e_{32} + 2 e_{42} - 3 d e_{42}) g_{20*}) \nonumber \\
&+& d (4 + d) ((-4 + 6 e_{32} + 3 d^2 e_{32} + 6 e_{42} + d
(-2 + 12 e_{32} + 3 e_{42}))
          g_{10*}^2 \nonumber \\
&-& (10 + 4 e_{12} + 2 d^2 e_{12} + 4 e_{22} + 6 e_{32} - 18 e_{42}
\nonumber \\
&+& d (-1 + 8 e_{12} + 2 e_{22} - 6 e_{32} + 3 e_{42})) g_{10*} g_{20*}
\nonumber \\
&+& ((2 + d + d^2) e_{12} + (-10 + d) e_{22} -
            3 (4 e_{32} + 2 d e_{32} + d^2 e_{32} \nonumber \\
&-& 4 e_{42} + 2 d e_{42})) g_{20*}^2)
       (-8 (2 + d) \epsilon + 3 (-1 + d)^2 (1 + d) (2 g_{10*} +
g_{20*}))))\,, \nonumber \\
g_{22d}&=&g_{21d}\,, \nonumber \\
e_{11n}&=& (g_q - g_s p_2) (g_p g_s (m_4 n_2 - m_3 n_3) p_1
\nonumber \\ &+&
+ g_{10*} g_o ( (m_4 n_1 + m_1 n_3) p_4 - (m_3 n_1 + m_1 n_2)
p_5))\,, \nonumber \\
 e_{d}&=& g_s^3 (m_4 n_2 - m_3 n_3) p_1 p_2 +
 g_{20*} g_o g_q
   (-(m_4 n_1 + m_1 n_3) p_4 + (m_3 n_1 + m_1 n_2) p_5) \nonumber \\
 &+& g_{20*} g_o g_s ((m_1 n_3 p_2 + m_4 n_1 (-p_1 + p_2)) p_4 +
      m_3 n_1 p_1 p_5 - m_3 n_1 p_2 p_5 \nonumber \\
&-& m_1 n_2 p_2 p_5 + m_2 p_1 (n_3 p_4 - n_2 p_5)) +
  g_k (g_q g_s (-(m_4 n_2) + m_3 n_3) \nonumber \\
&+& g_{20*} g_o (m_4 n_1 p_4 - m_2 n_3 p_4 - m_3 n_1 p_5 + m_2 n_2
p_5))\,, \nonumber \\
e_{12n}&=& (g_q - g_s p_2) (g_p g_s (m_4 n_2
          - m_3 n_3) p_2
\nonumber \\ &+&
+ g_{10*} g_o (- m_4 n_1 p_4 + m_2 n_3 p_4 + m_3 n_1 p_5 - m_2 n_2
p_5)) \,, \nonumber \\
 e_{21n}&=& (g_k - g_s p_1) (g_p g_s (m_4
n_2 - m_3 n_3) p_1
\nonumber \\ &+&
+ g_{10*} g_o (m_4 n_1 p_4 + m_1 n_3 p_4 - m_3 n_1 p_5 - m_1 n_2
p_5))\,, \nonumber \\
 e_{22n}&=& (g_k - g_s p_1) (g_p g_s (m_4
n_2 - m_3 n_3) p_2 \nonumber \\
&+& g_{10*} g_o (- m_4 n_1 p_4 + m_2 n_3 p_4 + m_3 n_1 p_5 - m_2 n_2
    p_5))\,, \nonumber \\
 e_{31n}&=& g_{20*} g_p g_s p_1 (-g_k m_4 n_1 + g_q m_4 n_1 +
g_q m_1 n_3 + g_k m_2 n_3 + g_s m_4 n_1 p_1 \nonumber \\
 &-& g_s m_2 n_3 p_1 - g_s m_4 n_1 p_2 - g_s m_1 n_3 p_2) +
  g_{10*} (g_s p_1 (g_s^2 (m_4 n_1 + m_1 n_3) p_2 \nonumber \\
  &-& g_{20*} g_o (m_1 + m_2) n_1 p_5) +
     g_k (-(g_q g_s (m_4 n_1 + m_1 n_3)) \nonumber \\
&+& g_{20*} g_o (m_1 + m_2) n_1
     p_5))\,, \nonumber \\
e_{32n}&=& g_{20*} g_p g_s p_2 (-g_k m_4 n_1 + g_q m_4 n_1 + g_q m_1 n_3 +
  g_k m_2 n_3 + g_s m_4 n_1 p_1 \nonumber \\
  &-& g_s m_2 n_3 p_1 - g_s m_4 n_1 p_2 - g_s m_1 n_3 p_2) +
  g_{10*} (g_k g_q g_s (m_4 n_1 - m_2 n_3) \nonumber \\
  &+& g_s^3 (-m_4 n_1 + m_2 n_3) p_1 p_2 -
     g_{20*} g_o g_q (m_1 + m_2) n_1 p_5 \nonumber \\
&+& g_{20*} g_o g_s (m_1 + m_2) n_1 p_2
     p_5)\,, \nonumber \\
 e_{41n}&=& g_{20*} g_p g_s p_1 (-g_k m_3 n_1 + g_q m_3 n_1 + g_q m_1 n_2 +
  g_k m_2 n_2 + g_s m_3 n_1 p_1 \nonumber \\
  &-& g_s m_2 n_2 p_1 - g_s m_3 n_1 p_2 - g_s m_1 n_2 p_2) +
  g_{10*} (g_s p_1 (g_s^2 (m_3 n_1 + m_1 n_2) p_2 \nonumber \\
  &-& g_{20*} g_o (m_1 + m_2) n_1 p_4) +
     g_k (-(g_q g_s (m_3 n_1 + m_1 n_2)) \nonumber \\
&+& g_{20*} g_o (m_1 + m_2) n_1
     p_4))\,, \nonumber \\
 e_{42n}&=& g_{20*} g_p g_s p_2 (-g_k m_3 n_1 + g_q m_3 n_1 + g_q m_1 n_2 +
   g_k m_2 n_2 + g_s m_3 n_1 p_1 \nonumber \\
   &-& g_s m_2 n_2 p_1 - g_s m_3 n_1 p_2 - g_s m_1 n_2 p_2) +
  g_{10*} (g_k g_q g_s (m_3 n_1 - m_2 n_2) \nonumber \\
  &+& g_s^3 (-m_3 n_1 + m_2 n_2) p_1 p_2
- g_{20*} g_o g_q (m_1 + m_2) n_1 p_4 \nonumber \\
&+& g_{20*} g_o g_s (m_1 +
m_2) n_1 p_2
  p_4)\,, \nonumber
\end{eqnarray}
where
\begin{eqnarray}
  l_1 &=& 24 + 16 d - 22 d^2 - 16 d^3 - 2 d^4,\nonumber \\
  m_1 &=& 48 - 16 d - 52 d^2 + 16 d^3 + 4 d^4,\nonumber \\
  m_2 &=& -48 - 80 d + 60 d^2 + 96 d^3 - 10 d^4 - 16 d^5 - 2 d^6,
\nonumber \\
  m_3 &=& -48 + 112 d + 32 d^2 - 130 d^3 + 14 d^4 + 18 d^5 + 2 d^6,
\nonumber \\
  m_4 &=& 48 + 104 d - 62 d^2 - 127 d^3 + 11 d^4 + 23 d^5 + 3 d^6,
\nonumber \\
  n_1 &=& 48 + 56 d - 40 d^2 - 56 d^3 - 8 d^4,\nonumber \\
  n_2 &=& 48 + 104 d - 32 d^2 - 104 d^3 - 16 d^4,\nonumber \\
  n_3 &=& -48 + 16 d + 10 d^2 - 41 d^3 + 35 d^4 + 25 d^5 + 3 d^6,
\nonumber \\
  o_1 &=& 26 - 7 d - 27 d^2 + 7 d^3 + d^4,\nonumber \\
  o_2 &=& -12 + 12 d^2,\nonumber \\
  p_1 &=& -96 - 64 d + 88 d^2 + 64 d^3 + 8 d^4,\nonumber \\
  p_2 &=& -96 - 64 d + 124 d^2 + 82 d^3 - 26 d^4 - 18 d^5 - 2 d^6,
\nonumber \\
  p_3 &=& 96 + 40 d - 140 d^2 - 60 d^3 + 42 d^4 + 20 d^5 + 2 d^6,
\nonumber \\
  p_4 &=& 144 + 96 d - 132 d^2 - 96 d^3 - 12 d^4,\nonumber \\
  p_5 &=& 144 + 96 d - 186 d^2 - 123 d^3 + 39 d^4 + 27 d^5 + 3 d^6,
\nonumber \\
  p_6 &=& -24 d - 52 d^2 + 4 d^3 + 50 d^4 + 20 d^5 + 2 d^6,\nonumber \\
  p_7 &=& 96 + 16 d - 192 d^2 - 56 d^3 + 92 d^4 + 40 d^5 + 4 d^6,
\nonumber \\
  q_1 &=& 96 + 16 d - 156 d^2 - 38 d^3 + 58 d^4 + 22 d^5 + 2 d^6,
\nonumber \\
  r_1 &=& 24 - 4 d - 36 d^2 + 2 d^3 + 12 d^4 + 2 d^5,\nonumber \\
  r_2 &=& 12 + 2 d - 18 d^2 - 3 d^3 + 6 d^4 + d^5,\nonumber \\
  r_3 &=& 12 - 6 d - 18 d^2 + 5 d^3 + 6 d^4 + d^5, \nonumber \\
  g_s &=& g_{10*} + g_{20*},\nonumber \\
 g_p &=& g_{10*} + g_{10*}^2/g_{20*},\nonumber \\
  g_k&=&(g_{10*}^2 p_3)/g_{20*} + g_{20*} p_6 + g_{10*} p_7,
  \nonumber \\
  g_q &=& (-(d g_{20*}^2 l_1) + g_{10*} (g_{10*} p_3 +
          g_{20*} q_1))/g_{20*}, \nonumber \\
 g_o &=& g_s^2/g_{20*}.\nonumber
\end{eqnarray}

Stability of the fixed point is determined by the $(2\times 2)$
block of the stability matrix which corresponds to $\beta$-functions
of $\alpha_{5}$ and $\chi_3$. The eigenvalue which
responds for instability has the form:
\begin{eqnarray}
\lambda = \lambda_0 + \lambda_1 \alpha_1 + \lambda_2
\alpha_2\,,\nonumber
\end{eqnarray}
where
\begin{eqnarray}
\lambda_0&=&\frac{d g_{20*}(g_{10*}+g_{20*})o_1-\sqrt{t_1} +
g_{10*} g_{20*} r_1+ g_{10*}^2 r_2 + g_{20*}^2 r_3}{8 d (12 + 8
d^2 + d^2) g_{20*}}\,, \nonumber \\
 \lambda_1 &=&\frac{\lambda_{1n}}{\lambda_{d}}\,, \nonumber \\
 \lambda_{1n}&=& d g_{20*}^2 (g_{11*}+g_{21*}) o_1 \sqrt{t_1} +
 g_{20*}\left(-t_2+g_{11*}\sqrt{t_1} \left(g_{20*} r_1 + 2 g_{10*}
r_2\right)\right)\nonumber \\
 &+& g_{21*}\sqrt{t_1}\left(\sqrt{t_1}-g_{10*}^2 r_2 + g_{20*}^2
 r_3 \right)\,, \nonumber \\
 \lambda_{d}&=&
 8 d (12 + 8 d + d^2)g_{20*}^2\sqrt{t_1}\,, \nonumber \\
 \lambda_2 &=& \frac{\lambda_{2n}}{\lambda_{d}}\,, \nonumber\\
 \lambda_{2n}&=& d g_{20*}^2 (g_{12*}+g_{22*}) o_1 \sqrt{t_1} +
 g_{20*}\left(-t_3+g_{12*}\sqrt{t_1} \left(g_{20*} r_1 +
2 g_{10*} r_2\right)\right)\nonumber \\
 &+& g_{22*}\sqrt{t_1}\left(\sqrt{t_1}-g_{10*}^2 r_2 + g_{20*}^2
 r_3 \right)\,, \nonumber
\end{eqnarray}
with
\begin{eqnarray}
t_1 &=& d^2 g_{20*}^2 (g_{10*} + g_{20*})^2 (o_1^2 - 4 o_2^2) -
   2 d g_{20*} o_1 (g_{10*} + g_{20*}) \times \nonumber \\
&\times &(g_{10*} g_{20*} r_1 + g_{10*}^2 r_2 + g_{20*}^2 r_3)
   + (g_{10*} g_{20*} r_1 + g_{10*}^2 r_2 + g_{20*}^2
   r_3)^2\,,\nonumber \\
t_2 &=& 2 (d^2 g_{20*} (g_{10*} + g_{20*}) (g_{11*} g_{20*} +
   (g_{10*} + 2 g_{20*}) g_{21*}) (o_1^2 - 4 o_2^2) \nonumber \\
   &+& (g_{10*} g_{20*} r_1 + g_{10*}^2 r_2 + g_{20*}^2 r_3)
     (g_{11*} g_{20*} r_1 + g_{10*} g_{21*} r_1 +
2 g_{10*} g_{11*} r_2 \nonumber \\
     &+& 2 g_{20*} g_{21*} r_3) -
    d o1 (g_{10*}^3 g_{21*} r_2 + g_{10*}^2 g_{20*}
(3 g_{11*} r_2 + 2 g_{21*} (r_1 + r_2))\nonumber \\
    &+& g_{20*}^3 (4 g_{21*} r_3 + g_{11*} (r_1 + r_3)) +
       g_{10*} g_{20*}^2 (2 g_{11*} (r_1 + r_2) + 3 g_{21*} (r_1 +
       r_3))))\,, \nonumber \\
t_3 &=& 2 (d^2 g_{20*} (g_{10*} + g_{20*}) (g_{12*} g_{20*} +
  (g_{10*} + 2 g_{20*}) g_{22*}) (o_1^2 - 4 o_2^2) \nonumber \\
  &+& (g_{10*} g_{20*} r_1 + g_{10*}^2 r_2 + g_{20*}^2 r_3)
     (g_{12*} g_{20*} r_1 + g_{10*} g_{22*} r_1 +
2 g_{10*} g_{12*} r_2 \nonumber \\
     &+& 2 g_{20*} g_{22*} r_3) -
    d o_1 (g_{10*}^3 g_{22*} r_2 + g_{10*}^2 g_{20*}
(3 g_{12*} r_2 + 2 g_{22*} (r_1 + r_2)) \nonumber \\
    &+& g_{20*}^3 (4 g_{22*} r_3 + g_{12*} (r_1 + r_3)) +
       g_{10*} g_{20*}^2 (2 g_{12*} (r_1 + r_2) + 3 g_{22*} (r_1 +
       r_3))))\,. \nonumber
\end{eqnarray}
borderline dimension $d_c$ is defined  as a solution of the equation
\begin{eqnarray}
\lambda (d_c, \epsilon, \alpha_1, \alpha_2)=0\,. \nonumber
\end{eqnarray}

\end{document}